\documentclass[pra,twocolumn,floatfix,a4paper,superscriptaddress,longbibliography]{revtex4-1}
\usepackage{bbm,color,graphicx,amsmath,txfonts}

\usepackage{hyperref}

\newcommand{\rp}[1]{(\ref{#1})}

\newcommand{\abs}[1]{\left|{#1}\right|}

\newcommand{\av}[1]{\left\langle #1 \right\rangle}

\newcommand{\wt}[0]{\widetilde}

\newcommand{\al}[1]{^{(#1)}}
\newcommand{\da}{^\dagger}

\newcommand{\pt}[1]{\left( #1 \right)}
\newcommand{\pq}[1]{\left[ #1 \right]}
\newcommand{\pg}[1]{\left\{ #1 \right\}}

\newcommand{\bs}[1]{\boldsymbol #1}

\newcommand{\lpg}[1]{\left\{ #1 \right.}

\newcommand{\rpg}[1]{\left. #1 \right\}}
\newcommand{\ee}{{\rm e}}
\newcommand{\ii}{{\rm i}}
\newcommand{\dd}{{\rm d}}

\newcommand{\id}{\mathbbm{1}}

\newcommand{\nn}{{\nonumber}}

\newcommand{\mat}[2]{
                      \begin{array}{#1}
                       #2
                       \end{array}  }

\newcommand{\mm}[2]{ \pt{
                      \begin{array}{cc}
                       #1 \\
                       #2
                     \end{array}  }  }

\newcommand{\EE}{{\cal E}}

\newcommand{\PP}{{\cal P}}
\newcommand{\QQ}{{\cal Q}}

\definecolor{blue}{rgb}{0,0,0.8}

\definecolor{green}{rgb}{0,0.6,0}

\usepackage[normalem]{ulem}

\newcommand{\stkout}[1]{\ifmmode\textrm{\sout{\ensuremath{#1}}}\else\sout{#1}\fi}

\begin{document}

\title{
Nonreciprocal conversion between radio-frequency and optical photons with an optoelectromechanical system}

\author{Najmeh Eshaqi-Sani}
\affiliation{Physics Division, School of Science and Technology, University of Camerino, I-62032 Camerino (MC), Italy}

\author{Stefano Zippilli}
\affiliation{Physics Division, School of Science and Technology, University of Camerino, I-62032 Camerino (MC), Italy}

\author{David Vitali}
\affiliation{Physics Division, School of Science and Technology, University of Camerino, I-62032 Camerino (MC), Italy}
\affiliation{INFN, Sezione di Perugia, via A. Pascoli, I-06123 Perugia, Italy}
\affiliation{CNR-INO, L.go Enrico Fermi 6, I-50125 Firenze, Italy}

\begin{abstract}
Nonreciprocal systems breaking time-reversal symmetry are essential tools in modern quantum technologies enabling the suppression of unwanted reflected signals or extraneous noise entering through detection ports. Here we propose a scheme enabling nonreciprocal conversion between optical and radio-frequency (rf) photons using exclusively optomechanical and electromechanical interactions. The nonreciprocal transmission is obtained by interference of two dissipative pathways of transmission between the two electromagnetic modes established through two distinct intermediate mechanical modes. In our protocol, we apply a bichromatic drive to the cavity mode and a single-tone drive to the rf resonator, and use the relative phase between the drive tones to obtain nonreciprocity.
We show that perfect nonreciprocal transduction can be obtained in the limit of large cooperativity in both directions, from optical to rf and vice versa. We also study the transducer noise and show that mechanical thermal noise is always reflected back onto the isolated port. In the limit of large cooperativity, the input noise is instead transmitted in an unaltered way in the allowed direction; in particular one has only vacuum noise in the output rf port in the case of optical-to-rf conversion.
\end{abstract}

\date{\today}
\maketitle

\section{Introduction}

Reciprocity is the two-way symmetry of transmission of light (photon) or sound (phonon) between forward and backward paths and is a common useful property exploited in a plethora of devices. However, when the time-reversal symmetry or reciprocity is broken, one can have novel functionalities that attracted considerable attention in engineered photonic systems~\cite{Gallo,Fang,Koch,Sounas,Kamal1,Hafezi}.

In fact, nonreciprocal transmission and amplification of signals are useful in communication, signal processing and measurement, because in nonreciprocal systems unwanted signals or spurious modes can be suppressed, thereby protecting the system from interference with extraneous noise~\cite{Pozar}. Typically, nonreciprocal devices require an element breaking Lorentz reciprocal symmetry~\cite{Feynman,Jalas} such as a d.c. magnetic field, but this method typically require bulky elements which are hard to integrate and miniaturize. Therefore, there is a strong motivation to realize alternative and more flexible implementations of nonreciprocity~\cite{Verhagen}. 
Various nonreciprocal devices were proposed and realized including magnetic materials~\cite{Auld,Milano,Fay,Aplet,Shirasaki,Sato,Bi}, or Josephson nonlinearities~\cite{Sliwa,Lecocq}, using temporal modulation~\cite{Anderson,Lira,Yu,Estep,Kang,Peng}, physical rotation~\cite{Fleury}, chiral atomic states~\cite{Scheucher}, and the quantum Hall effect~\cite{Mahoney}.

Recently Ref.~\cite{Metelmann} showed that a general recipe for obtaining nonreciprocal transmission is balancing any given coherent interaction with a properly tuned collective dissipative process. This insight led to propose and implement nonreciprocity using optomechanical devices where these ingredients are available and controllable. Multi-mode optomechanical and electromechanical schemes were proposed to achieve nonreciprocity and directionality, with or without relying on the direct coherent coupling between the electromagnetic input and output modes~\cite{Xu2016,Tian,Peterson,Bernier,Alu,Barzanjeh,cinesiwgm,Ruesink,DMalz,Lepinay}.
Here, similarly to the approach used in Refs.~\cite{Xu2016,Peterson,Bernier} which does not require any direct interaction between electromagnetic modes, we consider a four-mode optoelectromechanical system composed of an optical cavity and an rf resonator, each coupled to two intermediate mechanical modes. Two distinct paths of transmission between the two electromagnetic modes through the two mechanical modes are established
and their relative phase forms the basis of nonreciprocity and directionality. Differently from Refs.~\cite{Peterson,Bernier} which demonstrated the scheme in the microwave regime, here we exploit the possibility of mechanical modes to couple to fields of disparate wavelength, and we show the possibility of nonreciprocal conversion between optical and rf photons.
A similar, optical-microwave, four-modes nonreciprocal conversion scheme was proposed in Ref.~\cite{Xu2016} which however considered a four-tone driving scheme, in which both the optical and the microwave cavity are bichromatically driven. Here we simplify such a scheme and we consider an rf resonator driven by a single tone. In an appropriate parameter regime where the rotating wave approximation (RWA) also is valid, the system effectively becomes nonreciprocal and the transmission between the cavity and rf resonator is directional.
In this way one can add also the additional feature of nonreciprocity to the variety of optoelectromechanical devices, which were proposed and demonstrated for the transduction of rf and microwave signals to the optical domain~\cite{Taylor,Regal1,Barzanjehold,Bochmann,Andrews,PolzikNat,Balram,
Vainsecher,Takeda,Moaddel,Higginbotham,
Simonsen,Forsch,Jiang,Fink,Han,Arnold,Chu,Lambert}.

The outline of the paper is as follows. In Sec.~\ref{model}, the system and its Hamiltonian are introduced. In Sec.~\ref{Langevin}, the dynamics of our model described by Langevin-Heisenberg equations is studied, and the effective linearized model of the interacting four bosonic modes is obtained. In Sec.~\ref{Nonreciprocity} we analyze analytically the possibility to achieve nonreciprocity with this system, and in Sec.~\ref{Sec:noise} we study its noise properties. Sec.~\ref{numerical} is devoted to the numerical analysis where we determine the conditions where nonreciprocal optical-rf conversion is achieved, while concluding remarks are given in Sec.~\ref{conclusions}.

\section{The system}
\label{model}
We consider a hybrid optoelectromechanical system composed of an optical cavity coupled by radiation pressure to a mechanical element able to sustain multiple vibrational modes, which is in turn capacitively coupled to an rf resonant LC circuit.
Focusing on the case when only two nearby vibrational modes are coupled to the optical and rf resonators, the total Hamiltonian of the system can be written as the sum of an optical, mechanical and electrical contribution respectively,
\begin{equation}
  \hat{H}=\hat{H}_{\rm opt}+\hat{H}_{\rm mech}+\hat{H}_{\rm LC}.
  \label{hin}
\end{equation}
In more detail
\begin{eqnarray}
\hat{H}_{\rm opt}&=& \hbar \omega_{c}({x}_{1},{x}_{2})\:\hat{a}_{1}^\dag \hat{a}_{1} \\
&+& \hbar[(E_{1} e^{-\, i(\omega_{L1}t -\phi_{11})}+E_{2}
e^{-\, i(\omega_{L2}t -\, \phi_{12})})\hat{a}_{1}^\dag +h.c.] ,
\end{eqnarray}
where we consider a specific cavity mode, described by the photon annihilation (creation) operator $\hat{a}_{1}$ ($\hat{a}^\dag_{1}$), with the usual bosonic commutation relations $[\hat{a}_{1},\hat{a}^\dag_{1}]=1$, and bichromatically driven at two frequencies $\omega_{L1}$ and $\omega_{L2}$, with corresponding driving rates given by $E_j = \sqrt{2\kappa_{\rm in} \mathcal{P}_j/\hbar\omega_{Lj}}$, with $\mathcal{P}_j$ the $j$-th tone power and $\kappa_{in}$ the cavity amplitude decay rate through the input port. The mechanical term is
\begin{equation}
\hat{H}_{\rm mech} = \sum_{j=1,2}\frac{\hat{p}_{j}^2}{2 m_{j}}+\frac{m_{j} \omega_{j}^{2} \hat{x}_{j}^2}{2}
\end{equation}
where each mechanical resonator has effective mass $m_{j}$ $(j=1,2)$, displacement operator $\hat{x}_{j}$ and conjugated momentum $\hat{p}_{j}$, with commutation relations $[\hat{x}_{i},\hat{p}_{j}]=i\hbar\delta_{ij}$. Finally the rf circuit term is
\begin{equation}
\hat{H}_{\rm LC} = \frac{\hat{\phi}^2}{2 L}+\frac{\hat{q}^2}{2 C({x}_{1},{x}_{2})}- \hat{q}V_{\rm AC}\cos(\omega_{X}t-\phi_{X}),
\end{equation}
where $L$ is the inductance of the rf resonator, the dynamical variables of the LC circuit are given by the total charge and flux operators $\hat{q}$ and $\hat{\phi}$ respectively, with commutation relation $[\hat{q},\hat{\phi}]=i\hbar$, and the rf resonator is driven by a \emph{single-tone} drive at frequency $\omega_{X}$ and with voltage amplitude $V_{\rm AC}$.

\begin{figure}[t!]
\includegraphics[angle=0, width=1\linewidth]{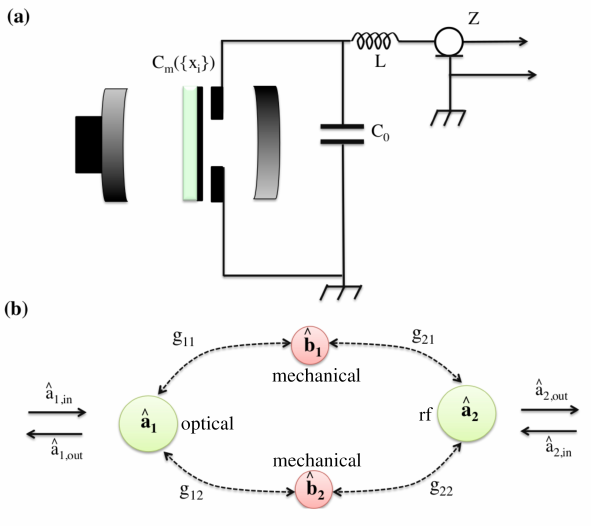}
\caption{
(a) Scheme of the proposed optoelectromechanical system. (b) Transmission pathways contributing to the non-reciprocal conversion.
The transmission from the optical input to the rf output or, vice versa, from the rf input to the optical output are mediated by the two mechanical resonators. These two transmission paths may interfere destructively in one direction but not in the opposite [depending on the phases and amplitudes of the complex interaction strengths $g_{\ell,j}$, see Eq.~\rp{GGGG}], hence realizing the non-reciprocal transduction.
}
\label{fig1}
\end{figure}

Such a configuration can be realized for example in the membrane-in-the-middle (MIM) optomechanical system case~\cite{HarrisNat,HarrisNJP,Kimble,Biancofiore,Karuza}, i.e., a driven optical Fabry-Per\'ot cavity with a thin semitransparent membrane inside. The membrane is metalized~\cite{Taylor,Andrews,PolzikNat,Moaddel,Regal2,Malossi} and capacitively coupled via an electrode to an LC resonant circuit formed by a coil and additional capacitors, see Fig.~\ref{fig1}.

The optomechanical and electromechanical couplings arise due to the dependence of the cavity mode frequency $\omega_{c}({x}_{1},{x}_{2})$ and of the circuit capacitance $C({x}_{1},{x}_{2})$ respectively, upon the displacement ${x}_{j}$ of the vibrational modes of the membrane. As in the scheme of Fig.~\ref{fig1}, the effective capacitance of the circuit is the parallel of a tunable capacitor $C_{0}$ with the membrane capacitor formed by the metalized membrane and an electrode in front of it, $C_{m}(\hat{x}_{1},\hat{x}_{2})$,
\begin{equation}
	C(\hat{x}_{1},\hat{x}_{2})=C_{0}+C_{m}(\hat{x}_{1},\hat{x}_{2})	.
\end{equation}
The system Hamiltonian of Eq.~(\ref{hin}) can be simplified by making two approximations: i) the two displacements ${x}_{j}$ are typically small and one can develop both the cavity frequency and the capacitance at first order in ${x}_{j}$; ii) one can neglect fast oscillating terms in the LC circuit driving. Moreover one can rewrite Eq.~(\ref{hin}) in a more convenient form by introducing the phonon annihilation and creation operators $\hat{b}_{j}$ and $\hat{b}_{j}^{\dagger}$, $j=1,2$, such that
\begin{equation}
\hat{x}_{j}\equiv x_{zpf,j}\,
\pt{\hat{b}_{j}+\hat{b}_{j}^{\dagger}}\ ,
\end{equation}
\begin{equation}
\hat{p}_{j}\equiv p_{zpf,j}\frac{\hat{b}_{j}-\hat{b}_{j}^{\dagger}}{i}\ ,
\end{equation}
where $x_{zpf,j}\equiv \sqrt{\frac{\hbar}{2m_{j}\omega_{j}}}$ and $p_{zpf,j}\equiv m\omega_{j} x_{zpf,j}$;%
and the LC photon annihilation and creation operators $\hat{a}_{2}$ and $\hat{a}_{2}^{\dagger}$, such that
\begin{equation}
\hat{q}\equiv q_{zpf}\,
\pt{\hat{a}_{2}+\hat{a}_{2}^{\dagger}}\ ,
\end{equation}
\begin{equation}
\hat{\phi}\equiv \phi_{zpf}\frac{\hat{a}_{2}-\hat{a}_{2}^{\dagger}}{i}\ ,
\end{equation}
(where the rf resonant frequency of the LC circuit is defined as $\omega_{LC}^{(0)}=1/\sqrt{LC(0,0)}$, and $ q_{zpf}\equiv \sqrt{\frac{\hbar}{2\,L\,\omega_{LC}^{(0)}}}$ and $\phi_{zpf}\equiv \sqrt{\frac{L\,\hbar\omega_{LC}^{(0)}}{2}}$). In the reference frame, for the optical mode, rotating at the frequency halfway between the two driving tones, $\omega_L = (\omega_{L1}+\omega_{L2})/2$, one finally obtains
\begin{eqnarray} \label{Hfin}
\hat{H}&=& \hbar \Delta_{L}\hat{a}_{1}^\dag \hat{a}_{1}+\hbar\sum_{j=1,2}g_{0,1j}(\hat{b}_{j}+\hat{b}_{j}^{\dagger})\hat{a}_{1}^{\dagger}\hat{a}_{1}
\nn\\\nonumber
&+&\sum_{j=1,2}\hbar\omega_{j}\hat{b}_{j}^{\dagger} \hat{b}_{j} + \hbar\omega_{LC}^{(0)}\hat{a}_{2}^{\dagger} \hat{a}_{2}
\\\nonumber
&-&\hbar\sum_{j=1,2}g_{0,2j}(\hat{b}_{j}+\hat{b}^{\dagger}_{j})(\hat{a}_{2}+\hat{a}_{2}^{\dagger})^{2}
\\\nonumber
&+& \hbar[(\mathcal{E}_{1} e^{i\omega_{+}t}+\mathcal{E}_{2} e^{-i\omega_{+}t})\hat{a}_{1}^\dag +h.c.]
\nn\\&-&
\hbar\pt{{V'}^*\,e^{i\omega_{X}t}+V'\,e^{-i\omega_{X}t}}\pt{\hat a_2+\hat a_2\da}
\end{eqnarray}
where we introduced the bare cavity detuning $\Delta_{L}\equiv \omega_{c}(0,0)-\omega_{L}$, $\omega_{+}\equiv\omega_{L1}-\omega_{L}=-(\omega_{L2}-\omega_{L})$, the single-photon optomechanical coupling rates $g_{0,1j}\equiv \frac{\partial\omega_{c}}{\partial x_{j}}|_{x_{i}=0}x_{zpf,j}$, the single-photon electromechanical coupling rates $g_{0,2j}\equiv \frac{\omega_{LC}^{(0)}}{4 C(0,0)}x_{zpf,j}\frac{\partial C}{\partial x_{j}}|_{x_{i}=0}$, the rf complex driving rate $V^{\prime}\equiv \left(q_{zpf}V_{\rm AC}/2\hbar
\right) e^{i\phi_{X}}$, and the complex optical driving rates $\mathcal{E}_{1}\equiv E_{1} e^{i\phi_{11}}$ and $\mathcal{E}_{2}\equiv E_{2} e^{i\phi_{12}}$.

\begin{figure}[!t]
\includegraphics[angle=0, width=1.0\linewidth]{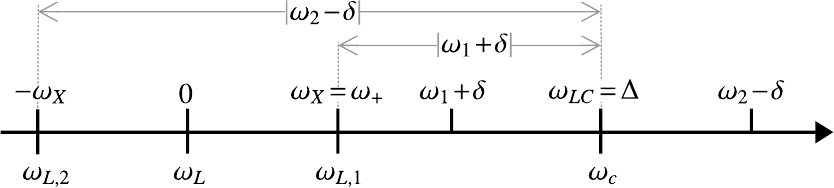}
\caption{
Frequency configuration. Above the horizontal axis the radio-frequency scale. Below the axis the optical scale.
}
\label{fig2}
\end{figure}
	
\section{Approximated model}\label{Langevin}

We now derive the quantum Langevin equations for the system operators by supplementing the Heisenberg equations of motion stemming from Eq.~(\ref{Hfin}) with fluctuation and dissipation terms describing the coupling of the two mechanical modes and of the two electromagnetic cavity modes with their own independent environment. We assume the ideal situation in which the optical cavity looses photons only from the input port with amplitude decay rate $\kappa_{in}\equiv \kappa$, and it is characterized by the input noise operator $\hat{a}_{1,in}$. We introduce damping and Brownian noise in a similar way for the two mechanical resonators, with energy decay rates $\gamma_{m,j}$ and noise operators
$\hat{b}_{j,in}$, $j=1,2$. For what concerns the LC circuit, we exploit the quantum electrical network theory of Ref.~\cite{Yurke} and model dissipation with an $RLC$ series circuit in which the input-output port is represented by an infinite transmission line with purely resistive characteristic impedance $Z=\sqrt{L_{T}/C_{T}}$, where $C_T$ and $L_T$ are the capacitance and the inductance per unit length along the transmission line, respectively. The input noise operator entering the circuit through the transmission line is $\hat{a}_{2,in}$. In an RLC series resonator the damping rate is $\gamma_{LC}\equiv Z/L$, and the rf-circuit quality factor is given by $Q_{LC}=\omega_{LC}^{(0)}/\gamma_{LC}$.

All the noise operators are uncorrelated from each other and characterized by thermal noise correlations at temperature $T$, where
the only non-zero correlation functions are
$\langle \hat{b}_{j,in}(t)\hat{b}_{j,in}^{\dagger}(t')\rangle = \langle \hat{b}_{j,in}^{\dagger}(t)\hat{b}_{j,in}(t')\rangle +\delta(t-t')=
\left[1+\bar{n}_{b\,j}\right]\delta(t-t')$, and $\langle \hat{a}_{j,in}(t)\hat{a}_{j,in}^{\dagger}(t')\rangle = \langle \hat{a}_{j,in}^{\dagger}(t)\hat{a}_{j,in}(t')\rangle +\delta(t-t')= \left[1+\tilde{n}_{a\,j}\right]\delta(t-t')$, with the number of thermal phonons given by $\bar{n}_{b\,j}=\left\{\exp[\hbar\omega_{j}/k_B T]-1\right\}^{-1}$, $j=1,2$, a similar expression for the mean thermal number of rf photons, $\tilde{n}_{a\,2}=\left\{\exp[\hbar\omega_{LC}^{(0)}/k_B T]-1\right\}^{-1}$, while $\tilde{n}_{a\,1}\simeq 0$ because at optical frequencies $\hbar\omega_{c} \gg k_B T$.

The quantum Langevin equations can then be written as
\begin{eqnarray}
\label{QLE0}
\dot{\hat{a}}_{1} &=& -(\kappa+i\Delta_{L})\,\hat{a}_{1}-i[\mathcal{E}_{1}e^{i\omega_{+}t}+\mathcal{E}_{2}e^{-i\omega_{+}t}]
\nn\\
&-&i\sum_{j=1,2}g_{0,1j}(\hat{b}_{j}+\hat{b}_{j}^{\dagger})\hat{a}_{1}
+\! \sqrt{2 \kappa}\hat{a}_{1,in}\!
\nn\\
\dot{\hat{a}}_{2} &=& -\left(\frac{\gamma_{LC}}{2}+i\omega_{LC}^{(0)}\right)\,\hat{a}_{2}+2i\sum_{j=1,2} g_{0,2j}(\hat{a}_{2}+\hat{a}_{2}^{\dagger})(\hat{b}_{j}+\hat{b}_{j}^{\dagger})
\nn\\
&+&i( {V^{\prime}}^* e^{i\omega_{X}t} +
 V^{\prime} e^{-i\omega_{X}t} )
+ \! \sqrt{\gamma_{LC}}\hat{a}_{2,in}\!
\nn\\
\dot{\hat{b}}_{j} &=& -\left(\frac{\gamma_{m,j}}{2}+i\omega_{j}\right)\,\hat{b}_{j}
-i g_{0,1j}\hat{a}_{1}^{\dagger}\hat{a}_{1}+i g_{0,2j}(\hat{a}_{2}+\hat{a}_{2}^{\dagger})^{2}
\nn\\
&+&\! \sqrt{\gamma_{m,j}}\hat{b}_{j,in}\!\ .
\end{eqnarray}

Here we are interested in the dynamics of the fluctuations $\delta\hat a_j=\hat a_j-\alpha_j(t)$ and $\delta\hat b_j=\hat b_j-\beta_j(t)$ about the corresponding mean amplitudes, $\alpha_j(t)=\av{\hat a_j}$ and $\beta_j(t)=\av{\hat b_j}$.
In order to study the corresponding dynamics we employ several approximations as detailed below (see also App.~\ref{App_approximations}).

\subsection{Linearization}

First, we linearize the equations for the fluctuations by assuming sufficiently large mean amplitudes.
In particular, we analyze the fluctuations in interaction picture with respect to the Hamiltonian
\begin{eqnarray}\label{H0}
\hat{H}_{0}&=&
\hbar\Delta\ \delta\hat{a}_{1}^{\dagger}\,\delta\hat{a}_{1}+\hbar\omega_{LC}\,\delta\hat{a}_{2}^{\dagger}\,\delta\hat{a}_{2}+\hbar\pt{\omega_{1}+\delta}\,\delta\hat{b}_{1}^{\dagger}\,\delta\hat{b}_{1}
\nn\\&&
+\hbar\pt{\omega_{2}-\delta}\,\delta\hat{b}_{2}^{\dagger}\,\delta\hat{b}_{2}\ ,
\end{eqnarray}
where $\Delta\equiv\Delta_{L}+2\sum_{j} g_{0,1j} Re\pg{\beta_{j}\al{dc}}$
and  $\omega_{LC}\equiv \omega_{LC}^{(0)}-4\sum_{j} g_{0,2j} Re\pg{\beta_{j}\al{dc}}$, with $\beta_j\al{dc}$ the time-independent part of the mean mechanical amplitude $\beta_j(t)$; And where $\delta$ is a small detuning which is used to tune the non-reciprocity as discussed in the following sections.
In this representation, the linearized quantum Langevin equations take the form
\begin{eqnarray}\label{QLEt}
\dot{\delta\hat{a}}_{1} &=& -\pq{\kappa+i\,\Theta_1(t)}\,\delta\hat a_1
\nn\\&&
-i\,\sum_{j=1,2}\pq{G_{1j}\al{-}(t)\ \delta\hat{b}_{j}+G_{1j}\al{+}(t)\ \delta\hat{b}_{j}^{\dagger}}
+ \sqrt{2 \kappa}\,\hat{a}_{1,in}
\nn\\
\dot{\delta\hat{a}}_{2} &=& -\pq{\frac{\gamma_{LC}}{2}+i\,\Theta_2(t)}\,\delta\hat{a}_{2}
+\Gamma(t)\ \delta\hat{a}_{2}\da
\nn\\&&
-i\,\sum_{j=1,2}\pq{G_{2j}\al{-}(t)\ \delta\hat{b}_{j}+G_{2j}\al{+}(t)\ \delta\hat{b}_{j}^{\dagger}}
+\sqrt{\gamma_{LC}}\,\hat{a}_{2,in}
\nn\\
\dot{\delta\hat{b}}_{j} &=& -\pq{\frac{\gamma_{m,j}}{2}+i(-1)^j\,\delta}\,\delta\hat{b}_{j}
\nn\\&&
-i\,\sum_{\ell=1,2}\pq{
G_{\ell j}\al{-}(t)^*\delta\hat{a}_\ell
+G_{\ell j}\al{+}(t)\,\delta\hat{a}_\ell^{\dagger}
}
+ \sqrt{\gamma_{m,j}}\,\hat{b}_{j,in}
\end{eqnarray}
where the time dependent coefficients can be expressed in terms of the mean field amplitudes $\alpha_j(t)$ and $\beta_j(t)$ as (see App.~\ref{App_approximations})
\begin{align}
\label{tpar}
\Theta_\ell(t)&=-(-2)^\ell\sum_{j=1,2} g_{0,\ell j}\ Re\pg{\beta_j(t)-\beta_{j}\al{dc}}
&{\rm for}\ \ell\in\pg{1,2}
\nn\\
G_{1j}\al{\pm}(t)&=g_{0,1j}\ \alpha_1(t)\
\ee^{i\,\pq{\Delta\pm\pt{\omega_j-(-1)^j\,\delta}}\,t}
&{\rm for}\ j\in\pg{1,2}
\nn\\
G_{2j}\al{\pm}(t)&=-4\,g_{0,2j}\ Re\pg{\alpha_2(t)}\
\ee^{i\,\pq{\omega_{LC}\pm\pt{\omega_j-(-1)^j\,\delta}}\,t}
&{\rm for}\ j\in\pg{1,2}
\nn\\
\Gamma(t)&=-4\,\sum_{j=1,2}g_{0,2j}\ Re\pg{\beta_j(t)}\ \ee^{i\,\omega_{LC}\,t}\ .
\end{align}
In particular, we remark that these parameters can be expanded as a sum of many terms each oscillating at a different frequency as
\begin{eqnarray}\label{Xomegaxi}
X(t)&=&\sum_\xi\ X_\xi\ \ee^{i\,\omega\al{X}_\xi\,t}
\hspace{0.6cm}{\rm for}\ \ X\in\pg{\Theta_\ell,G_{\ell,j}\al{\pm},\Gamma}\ ,
\end{eqnarray}
where the sum is over  all the possible frequency components, $\omega_\xi\al{X}$, of each parameter, and $X_\xi$ indicate the corresponding amplitudes (specific expressions for these quantities are reported in App.~\ref{App_approximations}, see also Ref.~\cite{Li}).

\subsection{Rotating wave approximation}

Then, we neglect all the time dependent terms. To be specific,
we note that when the system frequencies are selected such that (see Fig.~\ref{fig2})
\begin{eqnarray}\label{resonance}
\Delta-\omega_+&=&\omega_1+\delta
\nn\\
\Delta+\omega_+&=&\omega_2-\delta
\nn\\
\omega_{LC}-\omega_X&=&\omega_1+\delta
\nn\\
\omega_{LC}+\omega_X&=&\omega_2-\delta
\end{eqnarray}
which entail
\begin{eqnarray}\label{resonance0}
\Delta &=& \omega_{LC} = \frac{ \omega_2 + \omega_1 }{2}
\nn\\
\omega_+ &=& \omega_X = \frac{ \omega_2 - \omega_1-2\delta }{2} \ ,
\end{eqnarray}
only the terms $G_{\ell j}\al{-}(t)$ in Eq.~\rp{tpar} have a time independent part.
In the following we indicate the time independent part of
$G_{1 j}\al{-}(t)$ and of $G_{2 j}\al{-}(t)$ with the symbols
$g_{1 j}$ and $-g_{2 j}^*$ respectively (see App.~\ref{App_approximations} for details).
Moreover, all the remaining time-dependent terms can be neglected when (see App.~\ref{App_approximations})
\begin{eqnarray}\label{conditionsRWA}
\abs{g_{\ell j}}\ll\omega_1,\ \omega_2,\ \abs{\omega_1-\omega_2}\ .
\end{eqnarray}
Correspondingly, the quantum Langevin equation for the fluctuations reduce to the form
\begin{eqnarray}\label{QLE}
\dot{\delta\hat{a}}_{1} &=& -\kappa\,\delta\hat a_1
-i\,\sum_{j=1,2}\,g_{1j}\ \delta\hat{b}_{j}
+ \sqrt{2 \kappa}\,\hat{a}_{1,in}
\nn\\
\dot{\delta\hat{a}}_{2} &=& -\frac{\gamma_{LC}}{2}\,\delta\hat{a}_{2}
+i\,\sum_{j=1,2}\,g_{2j}^*\ \delta\hat{b}_{j}
+\sqrt{\gamma_{LC}}\,\hat{a}_{2,in}
\nn\\
\dot{\delta\hat{b}}_{j} &=& -\pq{\frac{\gamma_{m,j}}{2}+i(-1)^j\,\delta}\,\delta\hat{b}_{j}
-i\,g_{1 j}^*\,\delta\hat{a}_1
+i\,g_{2 j}\,\delta\hat{a}_2
\nn\\&&
+ \sqrt{\gamma_{m,j}}\,\hat{b}_{j,in}\ .
\end{eqnarray}
We also note that this equation can be valid only if the detuning $\delta$ is not too large, that is, it should be of the same order or smaller than the effective coupling coefficients
\begin{eqnarray}
\abs{\delta}\lesssim \abs{g_{\ell j}}\ .
\end{eqnarray}

\subsection{Perturbative expansion in powers of the bare couplings}

Finally, we compute explicit expressions for the interaction coefficients $g_{\ell j}$ by expanding the mean amplitudes $\alpha_j(t)$ and $\beta_j(t)$ in powers of the bare interaction coefficients $g_{0,\ell j}$. In particular, if the bare couplings are sufficiently small, then it is justifiable to consider only the corresponding leading zero-th order terms. In this way we find the following approximated expressions
\begin{eqnarray}\label{GGGG}
g_{1j}&=&-i\ g_{0,1j}\ \chi_1\ \EE_j
\nn\\
g_{21}&=&-4\ g_{0,21}\ Im\pg{\chi_{LC}}\, {V'}^*
\nn\\
g_{22}&=&-4\ g_{0,22}\ Im\pg{\chi_{LC}}\, {V'}
\end{eqnarray}
where we introduced the susceptibilities
\begin{eqnarray}
\chi_{1}&\equiv&\pq{\kappa+i\Delta_L}^{-1}\ ,
\nn\\
\chi_{LC}&\equiv&\pq{\frac{\gamma_{LC}}{2}+i\omega_{LC}^{(0)}}^{-1}\ .
\end{eqnarray}

Thereby we find an approximated model analogous to that of Refs.~\cite{Xu2016,Peterson,Bernier} which demonstrate non-reciprocity in a similar system where each mode of the electromagnetic field is driven by two pumps. Here we demonstrate the same behavior, when the low frequency mode (the rf-mode) is driven by a single pump. This is due to the fact that, when the frequency of the electromagnetic field is comparable to the mechanical frequencies, also the counter rotating terms in the pump can resonantly drive specific electromechanical processes.



\begin{figure*}[ht!]
\centering
\includegraphics[width=1\textwidth]{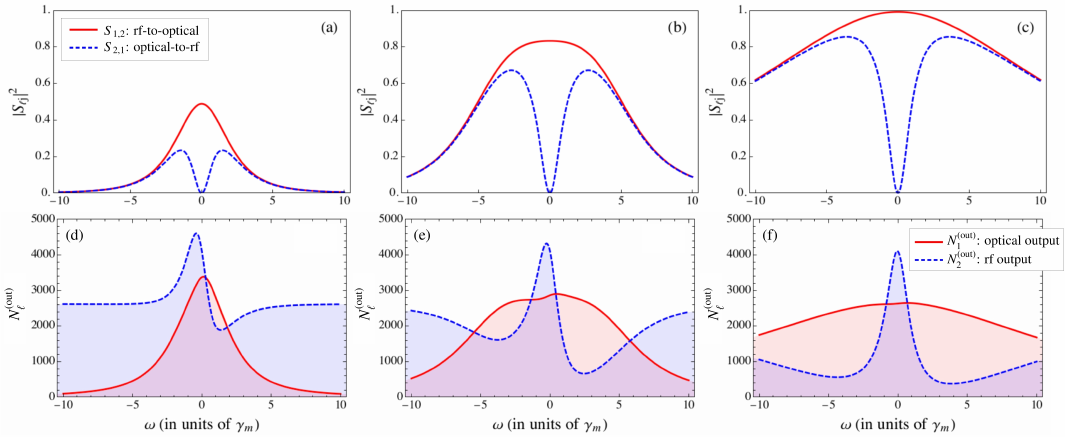}
\caption{
(a)-(c)
Transmission coefficients ($\abs{S_{21}}^2$, blue dashed lines, and $\abs{S_{12}}^2$, red solid lines), 
and (d)-(f) Noise spectral density of the fields at the output of the optical cavity (red solid line) and of the rf-resonator (blue dashed line),
when Eq.~\rp{c1} and \rp{c2} are true, the values of $E_{2}$ and $\varphi$ are set to fulfill Eqs.\eqref{varphir} and \eqref{opt}
for the suppression of the optical-to-rf transmission ($S_{21}=0$) at $\omega=0$, and $\abs{V'}$ and $\delta$ are set to fulfill Eqs.~\rp{c4} and \rp{c5} for the maximization of $S_{12}$.
In all plots $\omega_{m,1}=2~MHz$, $\omega_{m,2}=8~MHz$,
$\gamma_{LC}=6\,kHz$,
$\kappa=200\,kHz$,
$\gamma_m=500\,Hz$ and $g_0=3.5 Hz$.
In (a)
and (d)
$E_1=10\, GHz$ ($\Gamma=1$);
In (b) 
and (e)
$E_1=17.5\, GHz$ ($\Gamma=3$);
In (c) 
and (f)
$E_1=80\, GHz$ ($\Gamma=62.6$). 
In (d)-(f) the temperature is $0.1\,K$, corresponding to $\bar n_{a2}=2618$, $\bar n_{b1}=6545$ and $\bar n_{b2}=1636$.
In all the plots the other frequencies are fixed by the resonance conditions~\rp{resonance0}.
}
\label{fig3}
\end{figure*}
\begin{figure*}[ht!]
\centering
\includegraphics[width=1\textwidth]{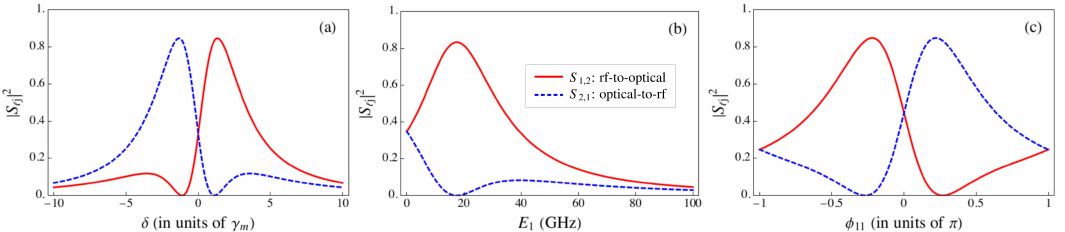}
\caption{
Transmission coefficients ($\abs{S_{21}}^2$, blue dashed lines, and $\abs{S_{12}}^2$, red solid lines) as a function of (a) the detuning $\delta$, (b) the strength of the first driving field $E_1$ and (c) the phase of the first driving field $\phi_{11}$, for $\omega=0$. The other parameters are as in Fig.~\ref{fig3} (b). The values of the parameters in the x-axis for which the dashed blue lines are zero are the values used in Fig.~\ref{fig3} (b). In (c) $\phi_{12}=\phi_X=0$, so that $\phi_{11}=\varphi$.
}
\label{fig4}
\end{figure*}

\begin{figure*}[t!]
\centering
\includegraphics[width=1\textwidth]{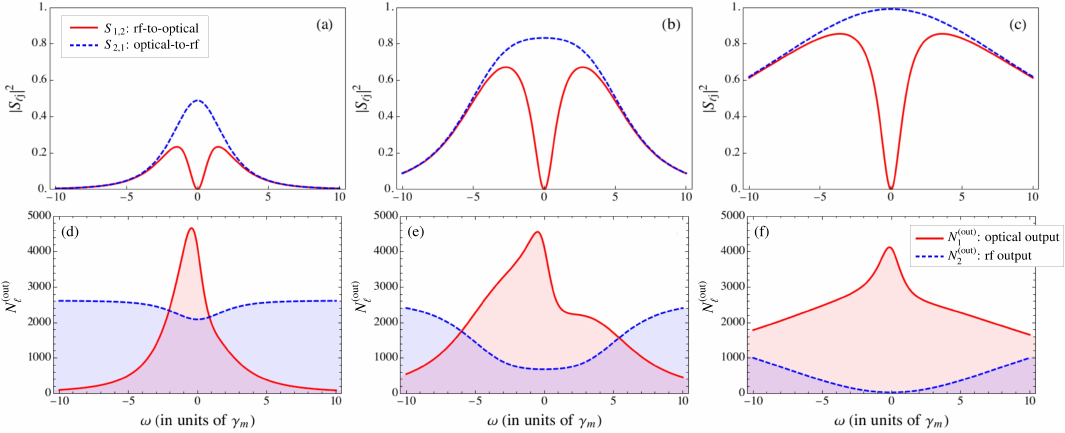}
\caption{
(a)-(c)
Transmission coefficients ($\abs{S_{21}}^2$, blue dashed lines, and $\abs{S_{12}}^2$, red solid lines), 
and (d)-(f) noise spectral density of the field at the output of the optical cavity (red solid line) and of the rf-resonator (blue dashed line),
when the values of $E_{2}$ and $\varphi$ are set to suppress the rf-to-optical transmission ($S_{12}=0$) at $\omega=0$, and $\abs{V'}$ and $\delta$ are set to maximize $S_{21}$.
The other parameters are as in Fig.~\ref{fig3}.
}
\label{fig6}
\end{figure*}

\section{Nonreciprocity}\label{Nonreciprocity}

In this section we study in detail the conditions for the non-reciprocal conversion, which can be derived from Eq.~\rp{QLE}, and we report results analogous to that discussed in Ref.~\cite{Peterson,Bernier}.
The equations~\rp{QLE} can be easily solved in Fourier space, and together with the standard input output relation
$\hat a_{1,out}=-\sqrt{2\,\kappa}\,\delta\hat a_1+\hat a_{1,in}$ and
$\hat a_{2,out}=-\sqrt{\gamma_{LC}}\,\delta\hat a_2+\hat a_{2,in}$, it is possible to express the output operators in terms of the input ones (see App.~\ref{App_Fourier}). In general, each output operator can be expanded as
\begin{eqnarray}\label{aout}
\hat a_{\ell,out}(\omega)=\sum_{j=1,2}\pq{ S_{\ell j}\ \hat a_{j,in}(\omega)+T_{\ell j}\ \hat b_{j,in}(\omega) }\ .
\end{eqnarray}
Here we are interested in the coefficients $S_{1 2}$ and $S_{2 1}$, which describe, respectively, how a radio-frequency input signal is converted into an optical field, and conversely how an optical signal is converted into a radio-frequency field.
Nonreciprocity corresponds to the situation in which one of these two coefficients is zero while the other is finite.
In general these quantities take the form
\begin{eqnarray}\label{Sjk}
S_{\ell j}=-\sqrt{2\,\kappa\,\gamma_{LC}}\ \frac{F_{\ell j}}{D}\ ,
\end{eqnarray}
with
\begin{eqnarray}\label{F12}
F_{12}&=&g_{11}\,\chi_{m,1}\,  g_{21}+g_{12}\, \chi_{m,2}\,g_{22}	
\\
F_{21}&=&g_{11}^{*}\,\chi_{m,1}\,g_{21}^{*}+g_{12}^{*}\,\chi_{m,2}\,g_{22}^{*}
\label{F21}
\end{eqnarray}
and
\begin{eqnarray}\label{D}
D&=&
\pt{\abs{g_{11}^2}\,\chi_{m,1}+\abs{g_{12}^2}\,\chi_{m,2}+\,\kappa-i\,\omega}
\\&&\times
\pt{\abs{g_{21}^2}\,\chi_{m,1}+\abs{g_{22}^2}\,\chi_{m,2}+\,\frac{\gamma_{LC}}{2}-i\,\omega}
-F_{12}\,F_{21} \ ,
\nn
\end{eqnarray}
and where we introduced the mechanical susceptibility in interaction picture
\begin{eqnarray}\label{chiM}
\chi_{m,j}=\pg{\frac{\gamma_{m,j}}{2}+i\pq{(-1)^j\,\delta-\omega}}^{-1}\ .
\end{eqnarray}
Eqs.~\rp{F12} and \rp{F21} indicate that each coefficient is the result of the interference of two transmission processes mediated by the two mechanical resonators. Here we look for situations in which the interference is destructive in one direction but not in the other.

We also note that $S_{21}$ is equal to the complex conjugate of $S_{12}$ evaluated at the opposite values of $\omega$ and $\delta$, i.e.
as functions of the frequency $\omega$ and of the detuning $\delta$
these coefficients fulfill the relation
\begin{eqnarray}\label{S12toS21}
S_{21}\pt{\omega,\delta}=
S_{12}^*\pt{-\omega,-\delta}\ .
\end{eqnarray}
This relation shows that to have nonreciprocity at $\omega=0$%
~\cite{footnote}, it is necessary to have $\delta\neq0$, otherwise the transmission in the two directions is necessarily symmetric.

If we introduce the parameters
\begin{eqnarray}\label{varphir}
\varphi&=&\phi_{11}-\phi_{12}-2\,\phi_X
\nn\\
r&=&\frac{g_{0,12}\,g_{0,22}}{g_{0,11}\,g_{0,21}}\ \frac{E_2}{E_1}
\end{eqnarray}
we find that $S_{21}=0$
when
\begin{eqnarray}\label{opt}
\ee^{-i\,\varphi}&=&-\abs{\frac{\chi_{m,1}}{\chi_{m,2}}}\,\frac{\chi_{m,2}}{\chi_{m,1}}
\nn\\
r&=&\abs{\frac{\chi_{m,1}}{\chi_{m,2}}}=\sqrt{\frac{\gamma_{m,2}^2+4\pt{\omega-\delta}^2}{\gamma_{m,1}^2+4\pt{\omega+\delta}^2}}\ .
\end{eqnarray}
Similarly $S_{12}=0$ when
$\ee^{i\,\varphi}=-r\ {\chi_{m,2}}/{\chi_{m,1}}$.

Let us now study the conditions under which $S_{12}$ achieves its maximum value when $S_{21}=0$. This can be done analytically when the two mechanical dissipation rates are equal
\begin{eqnarray}\label{c1}
\gamma_{m,j}\equiv\gamma_m\ \ \ \ {\rm for}\ j\in\pg{1,2}
\end{eqnarray}
and all the bare couplings $g_{0,\ell j}$ are equal, i.e.
\begin{eqnarray}\label{c2}
g_{0,\ell j}\equiv g_0\ \ \ \ {\rm for}\ \ell,j\in\pg{1,2}\ .
\end{eqnarray}
Moreover we assume the suppression of the transmission from optical to rf ($S_{21}=0$) at~\cite{footnote}
\begin{eqnarray}\label{c3}
\omega=0\ .
\end{eqnarray}
In this case $r=1$ [see Eq.~\rp{opt}], which entails $E_2=E_1$ [see Eq.~\rp{varphir}]. Thereby the transmission coefficient
$\abs{S_{12}}^2$ can be written in terms of the cooperativity parameters
\begin{eqnarray}\label{Gamma12}
\Gamma_1&=&\frac{2\,\abs{g_{1j}}^2}{\kappa\ \gamma_m}
\nn\\
\Gamma_2&=&\frac{4\,\abs{g_{2j}}^2}{\gamma_{LC}\ \gamma_m}
\end{eqnarray}
(note that in this case both $\abs{g_{11}}=\abs{g_{12}}$ and $\abs{g_{21}}=\abs{g_{22}}$)
as
\begin{eqnarray}
\abs{S_{12}}^2=\frac{
4\ \Gamma_1\ \Gamma_2\ \delta^2\ \gamma_m^4 \pt{\frac{\gamma_m^2}{4}+\delta^2}
}{
\pq{\pt{1+2\,\Gamma_1}\frac{\gamma_m^2}{4}+\delta^2}^2
\pq{\pt{1+2\,\Gamma_2}\frac{\gamma_m^2}{4}+\delta^2}^2
}\ .
\end{eqnarray}
The maximum of this expression is found for equal cooperativities
\begin{eqnarray}\label{c4}
\Gamma_j\equiv\Gamma=2\,\frac{\frac{\gamma_m^2}{4}+\delta^2}{\gamma_m^2} \ \ \ {\rm for}\ j\in\pg{1,2}
\end{eqnarray}
and it is given by $\abs{S_{12}}^2_{max}=\delta^2/\pt{\frac{\gamma_m^2}{4}+\delta^2}$.
The equality of the cooperativities can be realized by properly tuning the strength of the rf-pump $\abs{V'}$ [see Eqs.~\rp{GGGG} and \rp{Gamma12}]; Moreover, Eq.~\rp{c4} can be used to find the value of the detuning which maximizes $S_{12}$ when $S_{21}=0$,
\begin{eqnarray}\label{c5}
\delta=\pm\frac{\gamma_m}{2}\sqrt{2\,\Gamma-1}\ ,
\end{eqnarray}
and with this expression one can rewrite the maximum of the conversion coefficient as
\begin{eqnarray}\label{S12max}
\abs{S_{12}}^2_{max}=1-\frac{1}{2\,\Gamma}\ .
\end{eqnarray}
This expression shows that perfect conversion can be achieved in the limit of large cooperativity $\Gamma\to\infty$.

We also note that Eq.~\rp{S12toS21} can be used to find the analogous results corresponding to the suppression of $S_{12}$ and the corresponding maximization of $S_{21}$.
Finally, we point out that when the transmission in one direction is suppressed according to Eq.~\rp{opt} and in the other direction is maximized according to Eqs.~\rp{c1}-\rp{c3}, \rp{c4} and \rp{c5}, then also the reflection coefficients [$S_{11}$ and $S_{22}$ in Eq.~\rp{aout}] are suppressed [see Eqs.~\rp{ST} and \rp{ST2}]~\cite{Peterson,Bernier} as required by an isolator.

\section{
Output noise spectral density}
\label{Sec:noise}
So far we identified the conditions under which the transmission coefficient in one direction can be suppressed while the transmission in the opposite direction remains finite.
However, the full characterization of the capability of this system to be used as a nonreciprocal converter requires the study of how the noise associated to the various  components (in particular the mechanical and the rf noise) is redistributed in this system.
Here we analyze the noise spectral density at the output of both the optical cavity and of the rf-resonator. Differently from the results of Refs.~\cite{Peterson,Bernier}, here the two modes of the electromagnetic field have very different frequencies and the corresponding thermal noise is very different. Correspondingly, as shown below, the noise properties of the system are different when one suppresses either the rf-to-optical transmission ($S_{12}=0$) or the optical-to-rf transmission ($S_{21}=0$).

The symmetrized output noise spectral density is given by
\begin{eqnarray}\label{Nout}
N_\ell\al{out}(\omega)
&=&
\frac{1}{2}\int_{-\infty}^{+\infty}\dd\,t \
e^{i\,\omega\, t}
\av{
\hat{a}_{\ell,out}(t)\hat{a}_{\ell,out}^{\dagger}(0)
+
\hat{a}_{\ell,out}^{\dagger}(0)\hat{a}_{\ell,out}(t)
}
\nn\\
&=&
\frac{1}{2}
\abs{S_{\ell 1}(\omega)}^2+
\abs{S_{\ell 2}(\omega)}^2
\pt{\tilde{n}_{a2}+\frac{1}{2}}
\nn\\&&
+\abs{T_{\ell 1}(\omega)}^2\,\pt{\bar{n}_{b1}+\frac{1}{2}}
+\abs{T_{\ell 2}(\omega)}^2\,\pt{\bar{n}_{b2}+\frac{1}{2}}\ ,
\end{eqnarray}
where $S_{\ell j}$ and $T_{\ell j}$ were introduced in Eq.~\rp{aout} (and are explicitly defined in App.~\ref{App_Fourier}) and $\ell=1$ ($\ell=2$) is for the noise at the output of the optical cavity (rf-resonator).
%

When $S_{21}=0$ (suppression of optical-to-rf transmission) and $S_{12}$ (rf-to-optical conversion) is maximized  according to Eqs.~\rp{opt}-\rp{c3}, \rp{c4} and \rp{c5}, one finds (see App.~\ref{App_Fourier})
\begin{eqnarray}\label{Noutapprox}
N_1\al{out}(0)&=&
\pt{1-\frac{1}{2\,\Gamma}}
\pt{\tilde{n}_{a2}+\frac{1}{2}}
+\frac{1}{2\,\Gamma}\,
\pt{
\frac{\bar{n}_{b1}+\bar{n}_{b2}}{2}+\frac{1}{2}
}
\nn\\
N_2\al{out}(0)&=&
\frac{\bar{n}_{b1}+\bar{n}_{b2}}{2}+\frac{1}{2}\ ,
\end{eqnarray}
which show that in the limit  of large cooperativity $\Gamma\to\infty$ and around $\omega=0$, the rf-noise goes only into the optical output while the mechanical noise goes only into the rf-output.
This means that when this system is used, in this configuration, to convert a rf signal to the optical regime, the same number of thermal excitation of the rf field are also transferred to the optical output.
At the same time, the noise in the backward direction is increased because of the mechanical noise.
In fact, since the frequency of the rf resonator is equal to the average mechanical frequencies, see Eq.~\rp{resonance0}, then necessarily $\pt{\bar n_{b1}+\bar n_{b2}}/2>\bar n_{a2}$.

On the other hand, when $S_{12}=0$ (suppression of rf-to-optical transmission) and $S_{21}$ (optical-to-rf conversion) is maximized, one finds (see App.~\ref{App_Fourier})
\begin{eqnarray}\label{Noutapprox2}
N_1\al{out}(0)&=&\frac{\bar{n}_{b1}+\bar{n}_{b2}}{2}+\frac{1}{2}
\nn\\
N_2\al{out}(0)&=&
\frac{1}{2}\pt{1-\frac{1}{2\,\Gamma}}
+\frac{1}{2\,\Gamma}\,
\pt{
\frac{\bar{n}_{b1}+\bar{n}_{b2}}{2}+\frac{1}{2}
}\ .
\end{eqnarray}
Interestingly, in this case the contribution of the rf-noise in the output fields is completely suppressed around $\omega=0$. And, in the limit  of large cooperativity $\Gamma\to\infty$, the mechanical noise affects only the optical output such that the rf output reaches the vacuum noise level. Thus, in this limit the system realizes a quantum-limited optical-to-rf converter.


\section{Numerical results}\label{numerical}

We verified the non-reciprocity in this system numerically.
We studied both the rf-to-optical conversion with $S_{21}=0$ and the optical-to-rf conversion with $S_{12}=0$.

Fig.~\ref{fig3} corresponds to parameters for which the optical-to-rf transmission is suppressed ($S_{21}=0$), and the rf-to-optical conversion coefficient $S_{12}$ is maximized according to Eqs.~\rp{opt}-\rp{c3}, \rp{c4} and \rp{c5}. Plots from (a) to (c) correspond to increasing strength of the driving fields (i.e. increasing cooperativities). They show how, in agreement with Eq.~\rp{S12max}, the value of the conversion coefficient, at $\omega=0$~\cite{footnote}, increases with the cooperativity and approaches the value of 1 for large $\Gamma$.
These results are achieved by carefully tuning the driving strengths, frequencies and phases to their optimal working points.
In fig.~\rp{fig4} we varied the values of the system parameters so that the condition for the suppression of $S_{21}$, Eq.~\rp{opt}, and for the maximization of $S_{21}$, Eq.~\rp{c5}, are no longer satisfied (they are satisfied only in the points in which the dashed blue line is exactly zero).
This figure demonstrates that the non-reciprocal conversion is relatively robust to variation of the system parameters around their optimal values.

The output noise spectral density corresponding to Fig.~\ref{fig3} (a)-(c) is reported in Fig.~\ref{fig3} (d)-(f). As expected from the analysis of the previous section, both the optical and the rf outputs show increased noise around $\omega=0$. Moreover, at large frequency, the two output noise signals approach the level of noise of the free fields. Both the double peak structure, in the optical output, and the asymmetry of the plots are due to the mechanical noise:
The noise components associated to the two mechanical resonators are not peaked at $\omega=0$ because of the finite detuning $\delta$, and are asymmetric because of the different mechanical frequencies (which correspond to different numbers of thermal excitations).
We also note that by increasing the cooperativity, the width of the thermal peak increases in the optical output, and its maximum is reduced to the level of the rf noise.

These plots demonstrate the non-reciprocal conversion of an electromagnetic signal from the radio-frequency to the optical regime in an optoelectromechanical system which use only three driving fields.
In particular we observe that, for sufficiently large cooperativity, the conversion is perfect with $S_{12}=1$ and with a level of noise in the optical output which is equal to the noise of the rf input. However, while at the same time no optical signal can be transmitted to the rf-output (hence realizing the non-reciprocal conversion), significant mechanical noise reaches the rf-output. Thus the isolation of the rf port is not perfect. In principle this noise can be reduced using additional sideband cooling of the resonators~\cite{Bernier}.

The results corresponding to the suppression of the rf-to-optical transmission ($S_{12}=0$), and the maximization of the optical-to-rf conversion coefficient $S_{21}$,
are reported in Figs.~\ref{fig6}.
According to Eq.~\rp{S12toS21} these results are found by selecting the value of $\delta$, and correspondingly the value of $\varphi$ [see Eq.~\rp{opt}], opposite to those used in Figs.~\ref{fig3} [see Figs.~\ref{fig4} (a) and (c)]. We observe that the curves for the transmission coefficients in Fig.~\ref{fig6} are equal to those in Fig.~\ref{fig3} but with the exchanged role of $S_{12}$ and $S_{21}$. Moreover, the power spectral density of the output fields show that the noise of the rf field close to $\omega=0$ decreases with the strength of the driving fields, and as discussed in the previous section approaches the vacuum noise level at large cooperativity. However, at the same time, the mechanical noise is observable in the optical output.

\section{Conclusions}\label{conclusions}

In conclusion, we analyzed the possibility of achieving non-reciprocal transmission and conversion between optical and rf photons in an  optoelectromechanical system composed of an optical cavity, a rf LC-circuit and two mechanical resonators.

In this system the mechanical resonators mediate an indirect interaction between optical cavity and LC circuit, and the non-reciprocity relies on the interference between different transmission processes mediated by the two mechanical resonators and which result in different relative phases in the forward and backward directions.

We demonstrated that
non-reciprocity
is achievable also when only three fields (two optical and one rf) are used to drive the system.
This is possible because of the relatively small frequency of the rf mode, which is comparable to the frequencies of the mechanical resonators. In this case, counter rotating terms of the rf driving field may play the role of the fourth pump used in Refs.~\cite{Xu2016,Peterson,Bernier}.
Moreover we showed that, for sufficiently large cooperativity,  the non-reciprocal transduction is perfect in both directions, with the mechanical noise which affects only
the isolated port, and the input noise which is perfectly transmitted in the allowed direction. In particular, when the parameters are tuned to suppress the rf-to-optical transmission, this system can realize a quantum-limited optical-to-rf converter.

\begin{acknowledgments}
We acknowledge the support of the European Union Horizon 2020 Programme for Research and Innovation through the Project No. 862644 (FET Open QUARTET).
N.E.S acknowledges the TRIL support of the Abdus Salam International Centre of Theoretical Physics (ICTP).
\end{acknowledgments}

\appendix

\section{Approximations}\label{App_approximations}

The average amplitude of the electromagnetic and mechanical fields, $\alpha_j=\hat a_j-\delta\hat a_j$ and $\beta_j=\hat b_j-\delta\hat b_j$, fulfill the equations
[see Eq.~\rp{QLE0}]
\begin{eqnarray}\label{amplitudes}
\dot{\alpha}_{1}&=&-(\kappa+i\Delta_{L})\alpha_{1}-i[\mathcal{E}_{1}e^{i\omega_{+}t}+\mathcal{E}_{2}e^{-i\omega_{+}t}]
\\\nonumber
&-&i\sum_{j=1,2}g_{0,1j}	 \alpha_{1}(\beta_{j}+\beta_{j}^{*})
\\\nonumber
\dot{\alpha}_{2}&=&-(\frac{\gamma_{LC}}{2}+i\omega_{LC}^{(0)})\alpha_{2}
+2i\sum_{j=1,2}g_{0,2j}(\alpha_{2}+\alpha_{2}^{*})(\beta_{j}+\beta_{j}^{*})
\\
&+&
i\pq{V^{\prime} e^{-i\omega_{X}t}+{V^{\prime}}^* e^{i\omega_{X}t}}
\\
\dot{\beta}_{j}&=&-(\frac{\gamma_{m,j}}{2}+i\omega_{j})\beta_{j}-i g_{0,1j}|\alpha_{1}|^{2}\\\nonumber
&+&i g_{0,2j}(\alpha_{2}+\alpha_{2}^{*})^{2}\ .
\end{eqnarray}
The corresponding solutions
enter into the equations for the fluctuations, $\delta\hat a$ and $\delta\hat b$, as modulations of the interaction coefficients between different operators. When these rescaled interaction coefficients are sufficiently large, it is legitimate to linearize these equations, by neglecting non-linear terms in the fluctuations. In this way we find the linearized quantum Langevin equations for the fluctuations
\begin{eqnarray}
\dot{\delta\hat{a}}_{1} &=& -\pg{\kappa+i\,\pq{\Delta_L
+2\,\sum_{j=1,2} g_{0,1 j}\ Re\pg{\beta_j(t)}
}}\,\delta\hat a_1
\nn\\&&
-i\,\alpha_1(t)\sum_{j=1,2}g_{0,1j}
\pt{\delta\hat{b}_{j}+\delta\hat{b}_{j}^{\dagger}}
+ \sqrt{2 \kappa}\,\hat{a}_{1,in}
\nn\\
\dot{\delta\hat{a}}_{2} &=& -\pg{\frac{\gamma_{LC}}{2}+i\,\pq{\omega_{LC}\al{0}
-4\,\sum_{j=1,2} g_{0,2 j}\ Re\pg{\beta_j(t)}
}}\,\delta\hat{a}_{2}
\nn\\&&
+4\,i\sum_{j=1,2} g_{0,2j}\,Re\pg{\beta_j(t)}\ \delta\hat{a}_{2}\da
\nn\\&&
+4\,i\,Re\pg{\alpha_2(t)}\sum_{j=1,2} g_{0,2j}\,\pt{\delta\hat{b}_{j}+\delta\hat{b}_{j}^{\dagger}}
+\sqrt{\gamma_{LC}}\,\hat{a}_{2,in}
\nn\\
\dot{\delta\hat{b}}_{j} &=& -\left(\frac{\gamma_{m,j}}{2}+i\,\omega_{j}\right)\,\delta\hat{b}_{j}
-i g_{0,1j}\pq{\alpha_j(t)\,\delta\hat{a}_{1}^{\dagger}+\alpha_j(t)^*\delta\hat{a}_{1}}
\nn\\&&
+2\,i\, g_{0,2j}\,Re\pg{\alpha_2(t)}\pt{\hat{a}_{2}+\hat{a}_{2}^{\dagger}}
+ \sqrt{\gamma_{m,j}}\,\hat{b}_{j,in}
\end{eqnarray}
which are equivalent to the equations in interaction picture reported in the main text [see Eq.~\rp{QLEt} and \rp{tpar}].
As discussed in the main text, we evaluated explicit expressions for the coefficients~ \rp{tpar} by solving the equations for the mean amplitudes~\rp{amplitudes}. This can be done recursively by expanding the amplitudes in powers of the bare interaction coefficients $g_{0,ij}$.
When the interactions coefficients $g_{0,\ell j}$ are sufficiently small it is possible to consider only the first few terms of this expansion and neglect the rest. Here we consider coefficients up to their leading order~\cite{Li}, that is the zeroth order for $\alpha_j$ and the first order for $\beta_j$. We find that in the long time limit the mean amplitudes, $\alpha_j(t)$ and $\beta_j(t)$, are composed of a sum of terms which oscillate at multiples of the driving frequencies and at their sums end differences.

Specifically, we obtain the following expressions for the mean fields in the long time limit, up to the second order in $g_{0,ij}$
\begin{eqnarray}
\alpha_{1}(t)&\approx &
\alpha_{1,+}^{(0)}
\,\ee^{-i\,\omega_{+}t}
+\alpha_{1,-}^{(0)}
\,\ee^{i\,\omega_{+}t}
\nn	\\
\alpha_{2}(t)&\approx &
\alpha_{2,X}^{(0)}
\,\ee^{-i\,\omega_{X}t}
+\alpha_{2,-X}^{(0)}
\,\ee^{i\,\omega_{X}t}
\nn\\
\beta_{j}(t)&\approx &\beta_{j}^{dc}
+\beta_{j,+2}^{(1)}\,\ee^{-2\,i\,\omega_{+}t}
+\beta_{j,-2}^{(1)}\,\ee^{2\,i\,\omega_{+}t}
\nn\\&&
+\beta_{j,2X}^{(1)}\,\ee^{-2\,i\,\omega_{X}t}
+\beta_{j,-2X}^{(1)}\,\ee^{2\,i\,\omega_{X}t}	
\end{eqnarray}
where the zero-th order terms are
\begin{eqnarray}
\alpha_{1,+}^{(0)}&=& -i\,\chi_{1}\ \EE_1\ ,
\nn\\
\alpha_{1,-}^{(0)}&=& -i\,\chi_{1}\ \EE_2\ ,
\nn\\
\alpha_{2,X}^{(0)}&=& i\,\chi_{LC}\ V'\ ,
\nn\\
\alpha_{2,-X}^{(0)}
&=&
i\,\chi_{LC}\ {V'}^*\ ,
\end{eqnarray}
and the first order terms are
\begin{eqnarray}
\beta_{j}^{dc}&=&-i\,\chi^{\prime}_{m,j}~ \lpg{
g_{0,1j}
\pq{|\alpha_{1,+}^{(0)}|^{2}+|\alpha_{1,-}^{(0)}|^{2}}
}\nn\\&&
\rpg{
-2g_{0,2j}|\alpha_{2,X}^{(0)}+{\alpha_{2,-X}^{(0)}}^*|^{2}
}
\nn\\
\beta_{j,\pm\,2}^{(1)}&=&-i\,\chi_{m,j}^{\prime} ~g_{0,1j}~\alpha_{1,\pm}^{(0)}~\alpha_{1,\mp}^{(0) *}
\nn\\
\beta_{j,\pm 2X}^{(1)}&=&i\,\chi_{m,j}^{\prime} ~g_{0,2j}~\pq{\alpha_{2,\pm X}^{(0)}
+{\alpha_{2,\mp X}^{(0)}}^*}^{2}\ .
\end{eqnarray}

Correspondingly we find that the time dependent coefficients~\rp{tpar} of the linearized quantum
Langevin equations~\rp{QLEt} can be written as sums of many terms oscillating at different frequencies as in Eq.~\rp{Xomegaxi}.
To be specific, we find that the shifts of the electromagnetic frequencies can be written as
\begin{align}\label{Thetat}
\Theta_\ell(t)&=\sum_{\xi\in\pg{\pm}}\ \Theta_{\ell,\xi}\ \ee^{\ii\,\omega_{\xi}\al{\Theta}\,t}
&{\rm for}\ \ell\in\pg{1,2}
\end{align}
with frequencies
\begin{eqnarray}
\omega_{\xi}\al{\Theta}&=&\xi\ 2\,\omega_+\, \hspace{1cm}{\rm for}\ \xi\in\pg{\pm}
\end{eqnarray}
and corresponding coefficients
\begin{eqnarray}
\Theta_{\ell,\pm}&=&-(-2)^\ell\,\sum_{j=1,2}\ g_{0,\ell j}\,\pq{
\beta_{j,\pm 2}\al{1}+\beta_{j,\mp 2}^{(1)\,*}+\beta_{j,\pm 2X}\al{1}+\beta_{j,\mp 2X}^{(1)\,*}
}\ ;
\nn\\
\end{eqnarray}
The field enhanced interaction strengths are
\begin{eqnarray}\label{G12t}
G_{\ell,j}\al{\pm}(t)&=&
\sum_{\xi\in\pg{\pm}}\ G_{\ell,j,\xi}\al{\pm}
\ \ee^{i \omega_\xi\al{G\ell j\pm}\ t}
\hspace{0.5cm}{\rm for}\ \ell,j\in\pg{1,2}
\end{eqnarray}
with frequencies
\begin{eqnarray}
\omega_{\xi}\al{G1j\pm}&=&\Delta+\xi\pt{\omega_+-2\,\omega_X}\pm\wt\omega_j\, \hspace{1cm}{\rm for}\ \xi\in\pg{\pm}
\nn\\
\omega_{\xi}\al{G2j\pm}&=&\omega_{LC}-\xi\,\omega_X\pm\wt\omega_j\, \hspace{1cm}{\rm for}\ \xi\in\pg{\pm}
\end{eqnarray}
where
\begin{eqnarray}
\wt\omega_j=\omega_j-(-1)^j\,\delta\ ,
\end{eqnarray}
and corresponding coefficients
\begin{eqnarray}
G_{1,j,\xi}\al{\pm} &=&g_{0,1j}\ \alpha_{1,\xi}\al{0}
\nn\\
G_{2,j,\xi}\al{\pm} &=&-2\,g_{0,2j}\ \pq{
\alpha_{2,\xi X}\al{0} + \alpha_{2,-\xi X}^{(0)\ *}
}\ ;
\end{eqnarray}
And finally the self interaction strength of the rf mode, $\Gamma(t)$, is zero at this order of approximation.
We note that when the resonance conditions~\rp{resonance} are fulfilled, the frequencies
\begin{eqnarray}\label{freq}
\omega_{+}\al{G1 1-}&=&\Delta+\omega_+-2\,\omega_X-\wt\omega_1
\nn\\
\omega_{-}\al{G12-}&=&\Delta-\omega_++2\,\omega_X-\wt\omega_2
\nn\\
\omega_{+}\al{G21-}&=&\omega_{LC}-\omega_X-\wt\omega_1
\nn\\
\omega_{-}\al{G2 2-}&=&\omega_{LC}+\omega_X-\wt\omega_2
\end{eqnarray}
are zero. All the other frequencies, instead, are different form zero.
The frequencies in Eq.~\rp{freq} correspond, respectively, to the coefficients
\begin{eqnarray}
G_{1,1,+}\al{-}&=&g_{0,11}\ \alpha_{1,+}\al{0}
\nn\\
G_{1,2,-}\al{-}&=&g_{0,12}\ \alpha_{1,-}\al{0}
\nn\\
G_{2,1,+}\al{-}&=&-2\,g_{0,21}\ \pq{
\alpha_{2,X}\al{0} + \alpha_{2,-X}^{(0)\ *}
}
\nn\\
G_{2,2,-}\al{-}&=&-2\,g_{0,22}\ \pq{
\alpha_{2,-X}\al{0} + \alpha_{2,X}^{(0)\ *}
}\ .
\end{eqnarray}
In the main text we used the symbols $g_{\ell j}$ to indicate these coefficients, specifically, we used these definitions
\begin{eqnarray}
g_{11}&\equiv& G_{1,1,+}\al{-}
\nn\\
g_{12}&\equiv&G_{1,2,-}\al{-}
\nn\\
g_{21}&\equiv& -\,G_{2,1,+}^{(-)\,*}
\nn\\
g_{22}&\equiv& -\,G_{2,2,-}^{(-)\,*}\ ,
\end{eqnarray}
which are equal to the definitions in Eq.~\rp{GGGG}.
In our numerical simulations we have verified that all the other coefficients in Eqs.~\rp{Thetat}
and \rp{G12t} are much smaller than the corresponding frequencies, i.e $\abs{X_{\xi}}\ll\omega_{\xi}\al{X}$, for $X\in\pg{\Theta_\ell,G_{j,\ell}\al{\pm}}$ and for all corresponding $\ell$, $j$ and $\xi$.
In particular it is easy to check that these conditions are fulfilled when the conditions in Eq.~\rp{conditionsRWA} are true.

\section{The model in Fourier space}\label{App_Fourier}

Eq.~\rp{QLE} can be easily solved in Fourier space~\cite{Bernier}. To be specific one can express the mechanical operators in terms of the susceptibilities~\rp{chiM} and of the optical and rf mode operators as
\begin{eqnarray}
\delta\hat{b}_{1}(\omega) &=& \chi_{m,1}\ \big\{
   			-i g_{11}^{*}\hat{a}_{1}+i g_{21}\hat{a}_{2}+\! \sqrt{\gamma_{m,1}}\hat{b}_{1,in}\big\}
\nn\\   			
\delta\hat{b}_{2}(\omega) &=& \chi_{m,2}\ \big\{
   			-i g_{12}^{*}\hat{a}_{1}+i g_{22}\hat{a}_{2}+\! \sqrt{\gamma_{m,2}}\hat{b}_{2,in}	\big\}\ .
\end{eqnarray}
These expressions can be replaced into the equation for the electromagnetic fields and one obtain the following closed equation for the vector of operators $\boldsymbol{A}=(\delta\hat{a}_{1}(\omega),\delta\hat{a}_{2}(\omega))^{T}$,
\begin{eqnarray}
-i\,\omega\,\bs{A}=M\,\bs{A}+L\,\bs{A}_{in}+K\,\bs{B}_{in}
\end{eqnarray}
where $\boldsymbol{A}_{in}=(\hat{a}_{1,in}(\omega),\hat{a}_{2,in}(\omega))^{T}$, $\boldsymbol{B}_{in}=(\hat{b}_{1,in}(\omega),\hat{b}_{2,in}(\omega))^{T}$,
\begin{widetext}
\begin{eqnarray}
M=\left(\begin{array}{cccccc}
-(\kappa+|g_{11}|^{2}\chi_{m,1}+|g_{12}|^{2}\chi_{m,2}) &  g_{11}g_{21} \chi_{m,1}+g_{12}g_{22}\chi_{m,2} \\\nonumber
g_{11}^{*}g_{21}^{*} \chi_{m,1}+g_{12}^{*}g_{22}^{*}\chi_{m,2}& -(\frac{\gamma_{LC}}{2}+|g_{21}|^{2}\chi_{m,1}+|g_{22}|^{2}\chi_{m,2})\ , \\
\end{array}\right)
\end{eqnarray}
\end{widetext}
\begin{eqnarray}
L=\left(\begin{array}{cccccc}
\sqrt{2\kappa} &  0 \\\nonumber
0 & \sqrt{\gamma_{LC}} \\
\end{array}\right)\ ,
\end{eqnarray}
and
\begin{eqnarray}
K=\left(\begin{array}{cccccc}
-i g_{11}\chi_{m,1}\sqrt{\gamma_{m,1}} &  -i g_{12} \chi_{m,2}\sqrt{\gamma_{m,2}} \\\nonumber
i g_{21}^{*}\chi_{m,1}\sqrt{\gamma_{m,1}} & i g_{22}^{*}\chi_{m,2}\sqrt{\gamma}_{m,2} \\
\end{array}\right)\ ,
\end{eqnarray}
from which one finds the expressions of the modes operators in terms of the input noise operators
\begin{eqnarray}
\boldsymbol{A}=-[i\,\omega\,\id + M]^{-1}\,\pt{L~\boldsymbol{A}_{in}+K\,\boldsymbol{B}_{in}}\ .	
\end{eqnarray}
Finally using the input-output relations
\begin{eqnarray}
	\boldsymbol{A}_{out}=\boldsymbol{A}_{in}- L^{T}\boldsymbol{A}
\end{eqnarray}
with $\boldsymbol{A}_{out}=(\hat{a}_{1,out}(\omega),\hat{a}_{2,out}(\omega))^{T}$,
one finds the expression for the output fields~\rp{aout}
\begin{eqnarray}\label{Aout}
\bs{A}_{out}=S\ \bs{A}_{in}+ T\ \bs{B}_{in}
\end{eqnarray}
with
\begin{eqnarray}
S= \id + L^{T}\,[i\,\omega\, \id+ M]^{-1}\,L	
\end{eqnarray}
and
\begin{eqnarray}
T&=&
L^{T}\,[i\,\omega\, \id + M]^{-1}\,K
\nn\\
&=&
\pt{S-\id}\,L^{-1}\,K\ .
\end{eqnarray}
Specifically we find
\begin{eqnarray}
S&=&
\id-\frac{1}{D}\,L
\mm{F_{11}&F_{12}}{F_{21}&F_{22}}
\,L
\nn\\
&=&
\id-\frac{1}{D}\,
\mm{2\,\kappa\,F_{11}&\sqrt{2\,\kappa\,\gamma_{LC}}\,F_{12}}{\sqrt{2\,\kappa\,\gamma_{LC}}\,F_{21}&\gamma_{LC}\,F_{22}}
\end{eqnarray}
where $F_{12}$, $F_{21}$ and $D=F_{11}F_{22}-F_{12}F_{21}$ are defined in Eqs.~\rp{F12}-\rp{D} and
\begin{eqnarray}
F_{11}
&=&
\frac{\gamma_{LC}}{2}-i\,\omega+\abs{g_{21}}^2\,\chi_{m,1}+\abs{g_{22}}^2\,\chi_{m,2}
\nn\\
F_{22}
&=&
\kappa-i\,\omega+\abs{g_{11}}^2\,\chi_{m,1}+\abs{g_{12}}^2\,\chi_{m,2}\ ,
\end{eqnarray}
and
\begin{widetext}
\begin{eqnarray}
T&=&
-\frac{1}{D}\,L
\mm{F_{11}&F_{12}}{F_{21}&F_{22}}
\,K
\nn\\
&=&
\frac{i}{D}\,
\mm{
\sqrt{2\,\kappa\,\gamma_{m,1}}\ \chi_{m,1}\pq{
g_{11}\,F_{11}
-g_{21}^{*}\,F_{12}
}
&
\sqrt{2\,\kappa\,\gamma_{m,2}}\ \chi_{m,2}\pq{
g_{12}\,F_{11}
-g_{22}^{*}\,F_{12}
}
}{
\sqrt{\gamma_{LC}\,\gamma_{m,1}}\ \chi_{m,1}\pq{
g_{11}\,F_{21}
-g_{21}^{*}\,F_{22}
}
&
\sqrt{\gamma_{LC}\,\gamma_{m,2}}\ \chi_{m,2}\pq{
g_{12}\,F_{21}
-g_{22}^{*}\,F_{22}
}
}\ .
\end{eqnarray}
\end{widetext}
This expression shows that each coefficient of the matrix $T$, which describes the transfer of  mechanical noise to the electromagnatic fields (see Eq.~\rp{Aout}), is the sum of various terms which can interfere, and as discussed in Ref.~\cite{Bernier}, in same cases, certain terms can be suppressed.
In particular, when the parameters are chosen in order to suppress $S_{21}$ and to maximize $S_{12}$ according to Eqs.~\rp{opt}-\rp{c3}, \rp{c4} and \rp{c5}, one finds, for $\omega=0$ (see also Ref.~\cite{Peterson,Bernier})
\begin{eqnarray}\label{ST}
S&=&\PP\ \pt{\mat{cc}{
0 & \sqrt{1-\frac{1}{2\,\Gamma}}\\
0 & 0
}}
\nn\\
T&=&\frac{1}{\sqrt{2}}\ \PP\ \pt{\mat{cc}{
\frac{1}{\sqrt{2\,\Gamma}} & \frac{1}{\sqrt{2\,\Gamma}}\\
1 & -1
}}\ \QQ
\end{eqnarray}
with $\PP$ and $\QQ$ diagonal matrices, which include additional phases,
\begin{eqnarray}\label{PQ}
\PP&=&i\,\pt{\mat{cc}{
\ee^{i\pt{ \phi_\Delta+\phi_{12}+\phi_X\pm\phi_\Gamma }} & 0\\
0 & \pm 1
}}
\nn\\
\QQ&=&-i\,\pt{\mat{cc}{
\ee^{i\pt{\phi_X\pm\phi_\Gamma }} & 0\\
0 & \ee^{-i\pt{\phi_X\pm\phi_\Gamma }}
}}
\end{eqnarray}
where the sign $\pm$ corresponds to the sign in Eq.~\rp{c5}, and
\begin{eqnarray}
\ee^{i\,\phi_\Delta}&=&\frac{\kappa-i\,\Delta}{\sqrt{\kappa^2+\Delta^2}}
\nn\\
\ee^{i\,\phi_\Gamma}&=&\sqrt{1-\frac{1}{2\,\Gamma}}-\frac{i}{\sqrt{2\,\Gamma}}\ .
\end{eqnarray}
Eq.~\rp{ST} shows that in the limit of large cooperativity $\Gamma$, the optical output is not affected by the mechanical noise~\cite{Bernier}.
Similarly, when one suppresses $S_{12}$ and maximizes $S_{21}$, one finds
\begin{eqnarray}\label{ST2}
S&=&\wt\PP\ \pt{\mat{cc}{
0 & 0\\
\sqrt{1-\frac{1}{2\,\Gamma}} & 0
}}
\nn\\
T&=&\frac{1}{\sqrt{2}}\ \wt\PP\ \pt{\mat{cc}{
-1 & 1\\
\frac{1}{\sqrt{2\,\Gamma}} & \frac{1}{\sqrt{2\,\Gamma}}
}}\ \wt\QQ\ ,
\end{eqnarray}
with
\begin{eqnarray}\label{tildePQ}
\wt\PP&=&-i\,\pt{\mat{cc}{
\pm 1 & 0 \\
0 & \ee^{-i\pt{ \phi_\Delta+\phi_{12}+\phi_X\mp\phi_\Gamma }}
}}
\nn\\
\wt\QQ&=&\pt{\mat{cc}{
\ee^{i\pt{\phi_\Delta+\phi_{12}+2\,\phi_X\mp\phi_\Gamma }} & 0\\
0 & \ee^{i\pt{\phi_\Delta+\phi_{12}\mp\phi_\Gamma }}
}}\ .
\end{eqnarray}

The matrix $S$ in Eqs.~\rp{ST} and \rp{ST2} shows that the system behaves as a perfect isolator, where the transmission in one direction is large, and both the transmission in the other direction and the reflection coefficients are suppressed~\cite{Peterson,Bernier}.
Eqs.~\rp{ST} and \rp{ST2} can be used to compute the noise spectral density at the output of the optical and rf cavities [see Sec.~\ref{Sec:noise}]. In particular, we
note that the phases in Eqs.~\rp{PQ} and \rp{tildePQ} are irrelevant for this calculation, and one finds the expressions reported in Eqs.~\rp{Noutapprox} and \rp{Noutapprox2}.


\end{document}